\documentstyle[aps,multicol,psfig]{revtex}

\draft

\begin{document}

\title{Self-organization in populations of competing agents}

\author{Alexei V\'azquez$^{1,2}$}

\address{$^1$ Abdus Salam International Center for Theoretical Physics\\
        Strada Costiera 11, P.O. Box 586, 34100 Trieste, Italy}

\address{$^2$ Department of Theoretical Physics, Havana University\\
        San L\'azaro y L, Havana 10400, Cuba}

\maketitle

\begin{abstract}

A population of heterogenous agents compeeting through a minority rule is
investigated. Agents which frequently loose are selected for evolution by changing
their strategies. The stationary composition of the population resulting for this
self-organization process is computed analytically. Results are compared with
numerical simulations of two different minority games and other analytical
treatments available in the literature.

\end{abstract}

\begin{multicols}{2}\narrowtext

The mechanisms under which a population of compeeting agents self-organizes is a
problem which has gained a lot of imterest among the physics community in the last
few years \cite{arthur94,zhang97,johnson99,hulst99,ceva00,lo00,paczuski00}. As in
traditional statistical mechanism the main goal is to determine the average behavior
of the population based on the elementary rules which characterize the competition
among agents.  Given the actual state of art most of the works have been devoted to
the characterization, by means of numerical experiments or analytical treatments, of
the different models proposed in the literature
\cite{arthur94,zhang97,johnson99,paczuski00}, while a general description is still
missed. 

This work is an attempt to extract some general features present in the subclass of
models where the performance of the agents is characterized by certain minority rule
\cite{arthur94,zhang97,paczuski00}. Agents are assumed to be heterogeneous and the
main goal is to obtain which is the final composition of the population if certain
evolution mechanism is introduced.  The agents will be supposed not to be subscribed
to any regular lattice which rule out the existence of spatial correlations allowing
a mean field (MF) treatment. 

The work is organized as follows. First the general features of the class of models
under consideration are introduced. Based on these rules the MF rate equations
describing the evolution of the population are derived and the stationary solutions
are computed. These results are compared with numerical simulations of two
particular minority "games". Moreover, a comparison with the analytical results
already available for the model by Johnson {\em et al}
\cite{johnson99,hulst99,ceva00,lo00} is also drown. In all cases a very good
agreement is obtained. 

Considers a population of $N$ (odd) heterogeneous Boolean agents characterized by
certain property $x$. The Boolean nature of the agents restrict their action to only
two possibilities. An example is the Arthur-bar problem \cite{arthur94} in which
each agent attend or not a bar based on the past attendance to the bar. Another
example is the minority game introduced by Challet and Zhang \cite{zhang97} in which
the agents can either buy or sell. In both models a good choice of $x$ is the
probability that an agent takes different decisions in two different steps. A
different choice is taken in the model introduced by Johnson {\em et al}
\cite{johnson99}. In this case $x$ is the probability that an agent accepts the
decision suggested by its strategy or does the opposite. In general one make think
in $x$ as any property, or set of properties, which distinguish the different
strategies one agent can choose. 

On each step each agent take one of the two decisions and those being in the
minority win. The long time performance of the agents will be characterized by a
cumulative index $z$ in such a way that each time an agent win it is rewarded
with a positive point ($z\rightarrow z+1$), otherwise it is punished receiving a
negative point ($z\rightarrow z-1$). Evolution in the population is introduced
assuming that each time the performance $z$ of an agent goes below a threshold
$-z_c$ ($z_c>0$) it is selected for evolution an change its strategy: a new value
of $x$ ($x\rightarrow x^\prime$) is selected among certain distribution
$P_0(x^\prime)$ and its performance is reseted to zero ($z\rightarrow0$). 

A similar evolutionary mechanism has been already considered in \cite{johnson99}.
One also may think in other implementations like the extremal rules considered in
\cite{zhang97,paczuski00}, or a probabilistic rule in which agents can change
strategy even below $z_c$ according to certain probability, which in general may
depend on $z$. In any case it is expected that all of them give the same qualitative
behavior in the large $z_c$ limit.

Having defined the model let us determine which is the stationary composition of the
population of agents $P(x)$, the fraction of agents of type $x$.  For doing so we
first derive the rate equations which describe the dynamical evolution of $P(x)$,
which in general can be written as
\begin{equation}
\frac{\partial}{\partial t}P(x)=\sum_{x^\prime}\left[
W(x^\prime,x)P(x^\prime)-W(x,x^\prime)P(x)\right],
\label{eq:1} 
\end{equation} 
where $W(x^\prime,x)$ is the transition rate from $x^\prime$ to $x$.

The only rule which allows a change on $x$ is the evolution rule, in which the agent
chose a new value of $x^\prime$, selected with probability $P_0(x^\prime)$.  Thus,
if $\lambda(x)$ is the fraction of $x$-agents which are changing their strategy per
unit time then the transition rates are given by
$W(x,x^\prime)=\lambda(x)P_0(x^\prime)$. After substitution of this expression into
eq. (\ref{eq:1}) it results that
\begin{equation}
\frac{\partial}{\partial t}P(x)=P_0(x)\sum_{x^\prime}
\lambda(x^\prime)P(x^\prime)-\lambda(x)P(x).
\label{eq:2} 
\end{equation} 

To determine $\lambda(x)$ we need to consider the temporal evolution of the
cumulative performance index $z(x)$. Let $P_m(x)$ be the probability per unit
time that a $x$-agent is in the minority in one step. Thus, which probability
$P_m(x)$ ($1-P_m(x)$) the performance increases (decreases) by one unit. More
generaly one can define $P_m(x)$ as the probability that an agent wins a point,
regardless how it did. On the other hand, if an agent changes strategy, which
happens with probability $\lambda(x)$, then its performance increases by $z_c$.
These elementary processes lead to the rate equation
\begin{equation}
\frac{\partial}{\partial t} z(x)=2P_m(x)-1+z_c\lambda(x).
\label{eq:3}
\end{equation}

Provided $P_m(x)<1/2$ for all $x$ the system described by eqs. (\ref{eq:2}) and eq. 
(\ref{eq:3}) will always reach a stationary state. In this state time derivatives
vanish obtaining
\begin{equation}
\lambda(x)=\frac{1}{z_c}[1-2P_m(x)],
\label{eq:4}
\end{equation}
\begin{equation}
P(x)=\frac{A}{1-2P_m(x)}P_0(x),
\label{eq:5}
\end{equation}
where $A$ is a normalization constant. 

Agents with $P_m(x)$ close to $1/2$ are more stable to change in strategy as it can
be seen from eq. (\ref{eq:4}) and, therefore, they will have a larger participation
ratio in the stationary population, as follows from eq. (\ref{eq:5}). The
distribution from which the value of $x$ is extracted is thus modulated by the
probability of being in the minority, yielding the stationary population in eq.
(\ref{eq:5}).

On the other hand, from eq. (\ref{eq:4}) it can be seen that in the limit
$z_c\rightarrow\infty$ the fraction of agents changing strategy becomes
infinitesimal and, therefore, in this limit one expect to obtain the same results as
if one use an extremal evolution rule. Another aspect to be emphasized is the fact
that the parameter $z_c$, which in principle is the only evidence of the particular
evolution rule chosen here, does not appears in the expression for the stationary
distribution of agents in eq. (\ref{eq:5}) and, therefore, this result is expected
to hold independent of the particular evolution rule chosen. 

So far the results obtained here are very general and expected to apply to a wide
class of models of Boolean agents with minority rule. To go further we have to
determine $P_m(x)$ which may however depends on the particular model under
consideration. Below two different cases are analyzed. One is a very simple model of
non-interacting agents (NIA) which make their decision at random. The other is the
very recent implementation of a population of Boolean agents build up onto a
Kauffman's network (AKN)  introduced by Paczuski and Bassler \cite{paczuski00}. 

In the AKN model introduced in \cite{paczuski00} agents makes their decisions based
on the previous decision of some other agents. A Boolean variable $\sigma_i$
($i=1,2,\ldots,N$; $\sigma_i=0,1$) is assigned to each agent which is representative
of the two possible decisions one agent may take. The decision taken by each agent
is based on the decision took by other $K$ agents chosen at random
($i_1,\ldots,i_K$) in the previous step, according to certain Boolean function $f_i$
selected at random among the set of all the $2^{2^K}$ possible Boolean functions
with $K$ inputs, i.e.
\begin{equation}
\sigma_i(t+1)=f_i[\sigma_{i_1}(t),\ldots,\sigma_{i_K}(t)].
\label{eq:6}
\end{equation}
Without lost of generality and for a reason which becomes clear below, Booleans
functions which give the same output independent of the inputs are rule out.

In the original variant of this model \cite{paczuski00} an extremal evolution rule
is used, such that after certain time the worst agent is selected for evolution. 
This rule is relaxed here with the barrier evolution rule considered above, which
yields the extremal dynamics as $z_c\rightarrow\infty$. Moreover, one can see that a
property that makes differences among the agents can be the number of times 1
appears on the $M=2^K$ outputs of its Boolean function, denoted by $n$. If $n$ is
close to $M/2$ the agent will give as output 0 or 1, depending on the configuration
of its $K$ neighbors, with approximately the same probability.  Otherwise, if $n$
is close to 1 or $M$ the agent will practically give the same output independent of
the configuration of its neighbors. 

The NIA model is a simplification of the above model in which interactions among
agents are rule out. In this case on each step an $n$-agent gives 1 as output with a
probability $n/M$ and 0 otherwise. For this case the agent's decision does not
depends on decisions of any other agents and, therefore, there is no other
correlation than the one introduced by the minority rule which depends on the output
of all agents. 

Numerical simulations were performed for a population of $N=99$ agents and
$K=2,3,4,5$. In all cases the system is updated until it reaches the stationary
state and then average is taken over the temporal evolution of the population. The
resulting data is averaged over different realizations of the initial Boolean
functions and over the choice of neighbors in the interacting case.

Since the new Boolean functions are selected at random one has
\begin{equation}
P_0(n)=B(n;0.5,M)/\sum_{r=1}^{M-1}B(r;0.5,M),
\label{eq:6a}
\end{equation}
where $B(n;p,m)$ is the Binomial distribution, the probability to obtain 1 $n$ times
and $0$ $m-n$ times given on each single event 1 happens with probability $p$.  The
values $n=0$ and $M$ are rule out because they give fixed strategies and
consequently the Binomial distribution is renormalized. 

To compute the stationary composition $P(n)$ one has to compute the probability
$P_m(n)$ that an agent of type $n$ is in the minority, and then plug in the result
in eq. (\ref{eq:5}). For $N$ large the game is expected to be symmetric in the sense
that with probability $1/2$ the minority is the group of agents which takes the
1 (or 0) as output. In such a case a fixed agent will, in a long time window, have a
probability $1/2$ of being in the minority while agents changing their output very
often are expected to has a lower probability to be in the minority.

The main hypothesis taken here is that $1-2P_m(n)=f[\rho(n)]$, where $f[\rho(n)]$ is
a smooth function of the probability per unit time $\rho(n)$ that an agent change
its output. Since fixed players ($\rho=0$) has a probability $1/2$ to be in the
minority then $f(0)=0$. In the following $f(\rho)$ is expanded around $\rho=0$,
keeping only the linear term, resulting
\begin{equation}
1-2P_m(n)\propto \rho(n)+{\cal O}[\rho(n)^2].
\label{eq:7}
\end{equation}

Now, for NIA $\rho(n)$ is just the probability to find two different outputs on the
agent's strategy, which is given by $P_c(n)=2(n/M)(1-n/M)$. For AKN one should also
take into account that agents can changes their output only if at least one of its
inputs has changed output in the previous step. Within a MF approximation the last
event happens with probability $K\bar{\rho}$. Thus, in general
\begin{equation}
\rho(n)=BP_c(n),
\label{eq:8}
\end{equation}
where $B=1$ and $B=K\bar{\rho}$ for NIA and AKN, respectively.

By construction in both models $P_c(n)>0$ because the cases $n=0$ and $M$ have been
rule out. Hence, $1-2P_m(n)\propto BP_c(n)$ can only be zero if $B=0$ and if does it
is zero for all $n$. If the second possibility happens then all agents will have the
same probability to be in the minority and, therefore, their distribution in the
stationary state will be the distribution from where the Boolean functions are
extracted, i.e. $P(n)=P_0(n)$. Hence, eqs. (\ref{eq:5}),(\ref{eq:7})
and(\ref{eq:8}) yields the following alternative
\begin{equation}
P(n)=\left\{
\begin{array}{ll}
\frac{\bar{P_c}}{P_c(n)}P_0(n) & \text{for} B>0\\
P_0(n) & \text{for} B=0,
\end{array}
\right.
\label{eq:9}
\end{equation}
where $\bar{P_c}=\sum_nP_c(n)P(n)$, which is the final output of the present
calculation.

\begin{figure}
\centerline{\psfig{file=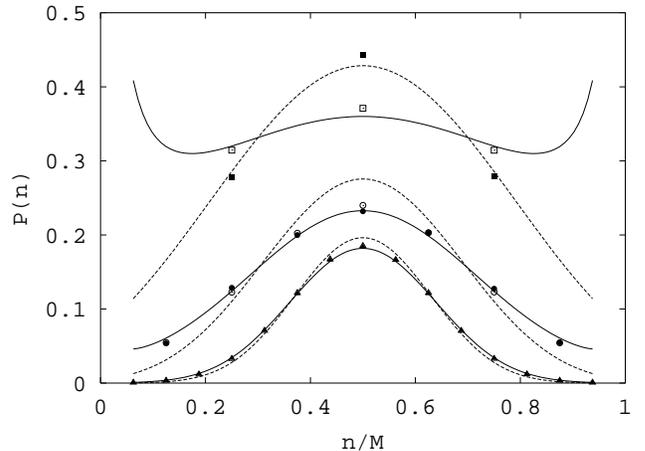,width=3.5in,angle=-90}}
\caption{Stationary composition of the population of agents. The points where
obtained from numerical simulations of a population of $N=99$ agents with a
threshold $z_c=100$ and $K=2$ (squares), $K=3$ (circles) and $K=4$ (triangles). In
all cases the open and full symbols corresponds to the NIA and AKN models,
respectively. The solid lines are obtained using eq. (\ref{eq:9}) with $B=0$
(dashed lines) and $B>0$ (solid lines) for the corresponding values of $K$. } 
\label{fig:1} 
\end{figure}

For NIA as mentioned above $B=1$ and, therefore, the alternative $B>0$ takes place.
The comparison of this prediction with numerical data is shown in fig. \ref{fig:1}.
The agreement is quite well for all values of $K$, proving that the ansatz in eq.
(\ref{eq:8}) applies for this model.

For AKN $B=K\bar{\rho}$ and one should analyze weather $\bar{\rho}$ is zero or not. 
As it is well known the Kauffman's network display qualitative different behavior
depending on the value of $K$ \cite{kauffman93}. For $K\leq2$ and independent of the
initial conditions the network evolve to a frozen configuration with
$\bar{\rho}=0$.  On the contrary for $K>2$ the network evolves to periodic orbits
with period growing exponentially with $N$ ("chaotic phase"), in which
$\bar{\rho}>0$.

Thus, the population of agents build up onto the Kauffman's network is a very good
scenario to test the validity of eq. (\ref{eq:9}) because both alternatives can be
observed. The comparison is shown in fig. \ref{fig:1}. For $K=2$ it can seen that
the numerical data is in better agreement by $P(n)=P_0(n)$ as predicted above. On
the contrary for $K>2$ the data is in better agreement with the case $B>0$ in eq. 
(\ref{eq:9}).

For $K>2$ both the NIA and AKN models yields the same distribution $P(n)$ and,
therefore, the correlations introduced by the network are in those cases irrelevant.
Moreover with increasing $K$ the composition of the population gradually approaches
$P_0(n)$ which explains the lost of self-organization observed in \cite{paczuski00}.

Finally the model of Johnson {\em et al} \cite{johnson99} is considered. In this
case the decisions taken by the agents depends on the previous history of the
winning group. For each agent it is available the information of which has been the
winning group $\sigma_w(t)$ ($\sigma_w(t)=0,1$) in the last $K$ steps. Moreover, to
each of then and strategy $f_i$is assigned, which is extracted at random among all
the possible Boolean functions of $K$ inputs. Then, for each possible past history
each agent will give a well defined output based on its own strategy, i.e. 
\begin{equation}
\sigma_i(t+1)=f_i[\sigma_W(t),\ldots,\sigma_W(t-K+1)].
\label{eq:10}
\end{equation}
Notices that in this case the output of each agent depends on a global information
and not on the outputs of any other agent as in the AKN model (see eq.
(\ref{eq:6})).

So far this model is just a variant of the minority game of Challet and Zhang
\cite{zhang97} for the case in which each agent has only one strategy at his
disposal. In the variation by Johnson {\em et al} \cite{johnson99} a probability
$p_i$ is assigned to each agent based on which the he accepts or not the output
of its strategy. With probability $p_i$ ($1-p_i$) he use the outcome (the opposite 
outcome) of its strategy. Moreover, they introduced the cumulative index $z_i$ to
measure the performance of each agent in a similar way as described above. The only
difference is that for this model when $z_i$ goes below the threshold $-z_c$ the
agent does not change its Boolean function $f_i$ but rather its probability $p_i$,
choosing a new one at random in the interval $[p_i-R/2,p_i+R/2]$ with reflective
boundary conditions.

Some analytical results are already available for this model
\cite{hulst99,lo00}. Using a diffusion like approach \cite{hulst99} or a detailed
probabilistic calculation \cite{lo00} it has been shown that in the stationary
population has the following composition
\begin{equation}
P(p)=\frac{A}{1-2P_m(p)},
\label{eq:11}
\end{equation}
where $A$ is a normalization constant and $P_m(p)$ is the probability that an agent
of type $p$ is in the minority \cite{note1}. 

Eq. (\ref{eq:11}) is actually quite similar to eq.  (\ref{eq:5}), with the choice
$x=p$. Since for this model the new values of $p$ are extracted from a uniform
distribution it is expected that $P_0(p)$ does not depend on $p$. Hence, eq.
(\ref{eq:11}) can be seen as a limiting case of eq. (\ref{eq:5}) when applied to
the particular evolution rule of the model by Johnson {\em et al}
\cite{johnson99} where $P_0(x)=const.$.

In order to go beyond this result one has to determine $P_m(p)$. This has already
been done in \cite{lo00} resulting that for large $N$ $1-2P_m(p)\approx C(N)2p(1-p)$
where $C(N)\sim N^{-1/2}$. This results does not seems to have any relation with the
ansatz in eq. (\ref{eq:9}). However, one should notice that $2p(1-p)$ is just the
probability $P_c(p)$ that an agent of type $p$ give two different outputs, given the
output of its Boolean function has remained fixed. Therefore, from this result
and eq. (\ref{eq:11}) it follows that
\begin{equation}
P(p)=\frac{\bar{P_c}}{P_c(p)}.
\label{eq:12}
\end{equation}
This eq. is just the $B>0$ alternative of eq. (\ref{eq:9}). Hence, the functional
dependence of $P_m(x)$ and $P(x)$ on $P_c(x)$ appears to be universal for the class
of models studied here.

In summary, a heterogenous populations of compeeting agents has been studied. Based
on general arguments, such as the minority rule and evolution, the stationary
composition of the population of agents has been computed as a function of the
probability $P_m(x)$ of being in the minority. Further analysis reveals that the
relation $1-2P_m(x)\propto P_c(x)$ is universal for the class of model, where
$P_c(x)$ is the probability that an agent of type $x$ gives different outputs.

For the AKN it is concluded that except for $K=2$ the correlations introduced by
the network are irrelevant and agents can be considered to be independent. In the
particular case $K=2$, which correspond to the critical network, the population
build-up onto the Kauffman's network reach an stationary state in which all
agents have a probability $1/2$ to be or not in the minority. Moreover, the lost of
self-organization with increasing $K$ was shown to takes place gradually.

\section*{Acknowledgements}

I thanks M. Paczuski for for useful comments and discussion. I also thanks Y.-C.
Zhang and M. Marsili for reading the manuscript.

\end{multicols}


\begin{thebibliography}{99}
\bibitem{note1} A different notation is used in \cite{hulst99,lo00} in which
$P_m(p)$ is written as $\tau(p)$.
\end{thebibliography}
\end{document}